# SimCleaner - Sistema de Padronização de Bases de Dados utilizando Funções de Similaridade

**Carlos Diego Nascimento Damasceno, Fabio Manoel França Lobato, Elton Rocha Moutinho, Arilene Santos de França, Ivan Ikikame de Oliveira, Ádamo Lima de Santana**

Laboratório de Planejamento de Redes de Alto Desempenho - Universidade Federal do Pará

Rua Augusto Corrêa, 01 – CEP 66075-110 – Belém – Pará - Brasil

```
{damasceno.diego,eltonmoutinho}@gmail.com,arilene_sf@msn.com,ikikameo@
                hotmail.com,{adamo,lobato.fabio}@ufpa.br
```

***Abstract.*** *The Knowledge Discovery in Database (KDD) process permits the detection of pattern in databases, where this analysis may be compromised if database is not consistent, making necessary the use of data cleaning techniques. This paper presents a tool based in similarity functions to help the preprocessing of databases and it behaved efficiently in the standardization of a System of Public Security of the State of Pará database and may be reused with other databases and other data mining projects.*

***Resumo.*** *O processo de descoberta de conhecimento em bases de dados (BD) permite detecção de padrões em BD, onde esta analise pode ser comprometida se o BD não estiver consistente, tornando necessário o uso de técnicas de data cleaning. Este artigo apresenta uma ferramenta baseada em funções de similaridade para auxiliar a fase de pré-processamento de dados, ela se mostrou eficiente na padronização de um BD do Sistema de Segurança Pública do Estado do Pará e pode ser reutilizada com outras bases e em outros projetos mineração de dados.*

## 1. Introdução

O processo de descoberta de conhecimento em bases de dados (BD) em inglês, *Knowledge Discovery in Database* (KDD) envolve basicamente três processos hierarquizados, seleção, pré-processamento e validação dos resultados, onde estas duas primeiras etapas demandam cerca de 80% do tempo total e são fases cruciais para o sucesso do processo [Winkler, 2004], pois é onde os dados são tratados, para que então possam ser submetidos às técnicas de inteligência computacional (IC).

Segundo [Lawrence et al. 1999], um problema muito enfrentado na integração de BD é o gerenciamento de dados duplicados; [Laender et al. 2004] ilustra essa problemática usando como exemplo o contexto encontrado no banco de dados BDBComp. O cadastramento de ocorrências policiais é outro exemplo deste contexto, sendo esta a problemática e motivação para o desenvolvimento deste trabalho.

Para solucionar este problema e assegurar a qualidade das informações, surgiu uma área denominada *data cleaning* que procura detectar e corrigir anormalidades em bases de dados. Essa detecção pode ser efetuada usando métodos estatísticos [Rham, 2000] ou de reconhecimento de padrões, como algoritmos de detecção de similaridade.

[Naumann, 2010] e [Dorneles, 2008] classificam as funções de similaridades como baseadas: em caractere, onde a análise é feita em cada caractere individualmente (Algoritmo de Levenshtein e Jaro Winkler), e baseada em *tokens*, que consiste na análise de cada *token* de uma *string* (algoritmo Jaccard.e Monge-Elkan).

A aplicação desenvolvida propõe a padronização de bases de dados usando as funções Levenshtein e Jaro-Winkler, onde o resultado das funções de similaridade é usado para associar termos equivalentes. Ela pode ser obtida em http://sourceforge.net/projects/simcleaner/.

## 2. Trabalhos Correlatos

Segundo [Chen 2005], cerca de 99% das 500 maiores empresas do mundo utilizam *data warehouse* para alimentar sistemas inteligentes de suporte a decisão, onde a consistência destes dados interfere na qualidade dos resultados obtidos.

Neste contexto, destaca-se o trabalho de [Yuet al. 2009], que propôs um *framework* universal para o processo de *Data Cleaning* baseado em modelos de usuários, capaz de extrair regras e armazená-las em arquivos XML (Extensible Markup Language), não chegando a implementá-lo, e o de [Christen 2009] que desenvolveu uma ferramenta para limpeza e integração, utilizando outros projetos *Open Source* como o *Febrl* (*Freely Extensible Biomedical Record Linkage*).

Entretanto, a proposta de [Christen 2009] é de uma padronização automática, algo que pode gerar falsos positivos, por exemplo, associando corretamente os nomes "Almirante Barroso, Alameda" e "Almirante Barroso, Avenida" por terem um percentual de similaridade de aproximadamente 88%, e não associando "Avenida Almirante Barroso" e "Almirante Barroso, Avenida" por ter apenas 66% de percentual de similaridade.

Para contornar isso, este trabalho propõe uma solução de funcionamento assistido, possuindo uma fase automática, onde associações automáticas são feitas, e outra manual, de validação das mudanças a serem realizadas.

## 3. Domínio de aplicação e estrutura de dados utilizada

O presente trabalho teve como estudo de casos uma BD de propriedade do Sistema de Segurança Pública do Estado do Pará (SSP-PA) relativa aos registros de ocorrências policiais da cidade de Belém do Pará, efetuados entre os anos de 2002 e 2009. Os registros se apresentavam de uma forma não homogênea, necessitando ser retificada para a posterior submissão ao processo de mineração de dados. O objetivo final seria o reconhecimento estatístico de padrões criminais.

A base possuía 963.951 registros de boletins de ocorrência (BO) da SSP-PA. Endereço do incidente, Tipo de crime e Nome do Logradouro são exemplos de alguns dos atributos da tabela. Após uma análise prévia, foram identificadas inconsistências que degradariam o processo de mineração de dados, destacando-se:

- Dados não padronizados: instâncias de um mesmo campo não seguiam um mesmo formato, por exemplo: "Bernardo Sayão, Avenida - de 2312/2313 a 3366/3367", "BERNARDO SAYÃO, AV.","BernardoSayão, AV.","Bernardo SAYÃO, AV.";

- Presença de *outliers (*valores que fogem ao padrão), como caracteres repetidos ("######", "XXXXX.");

Com os principais problemas identificados, foi iniciado o desenvolvimento de uma aplicação que solucionasse o problema de limpeza e padronização desta BD.

## 4. Especificação e Implementação da ferramenta

O desenvolvimento da solução para padronizar os campos da BD seguiu a hierarquia básica de desenvolvimento de software: extração de requisitos, projeto, implementação e testes. A seguir, estas etapas são brevemente descritas, juntamente com um esboço da arquitetura do *software,* de acordo com seu respectivo uso sobre o domínio.

### 4.1. Extração de Requisitos

Após a análise do problema, foi detectada a necessidade de uma solução que permitisse a validação das substituições antes de elas serem efetuadas, isso evitaria falsos positivos, e trataria dados ausentes e caracteres repetidos. A estratégia escolhida para solucionar o problema foi agrupar palavras similares em um dicionário e associá-las a um valor único, salvando em um arquivo. O dicionário deveria ser validado para que os possíveis erros fossem detectados e removidos, aumentando a precisão da filtragem.

### 4.2. Projeto e Implementação

Com a meta de facilitar a manipulação e o uso da aplicação em diversas plataformas foi escolhida a linguagem de programação Java, e a elaboração de uma interface gráfica amigável e intuitiva na qual todas as funcionalidades, como, por exemplo, a escolha do banco de dados e a edição do dicionário (**Figura 1 e Figura 2**) estivessem disponíveis de modo que o sistema pudesse ser operado facilmente pelo usuário.

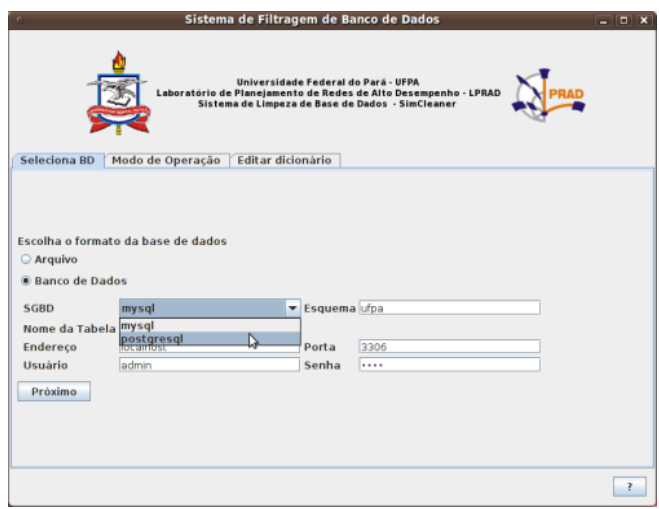

**Figura 1. Seleção de base de dados.**

O sistema suporta os formatos XLS e XLSX do *Microsoft™ Excel*, CSV (*Comma Separated Values*) e os SGBDs (Sistema de Gerenciamento de Banco de Dados) MySQL (mysql.com) e PostgreSQL (postgresql.org), sendo que essas compatibilidades foram inseridas com o uso de componentes, como as bibliotecas *opencsv* (opencsv.sourceforge.net), *Apache POI* (poi.apache.org) e *drivers* JDBC (*Java Database Connectivity*).

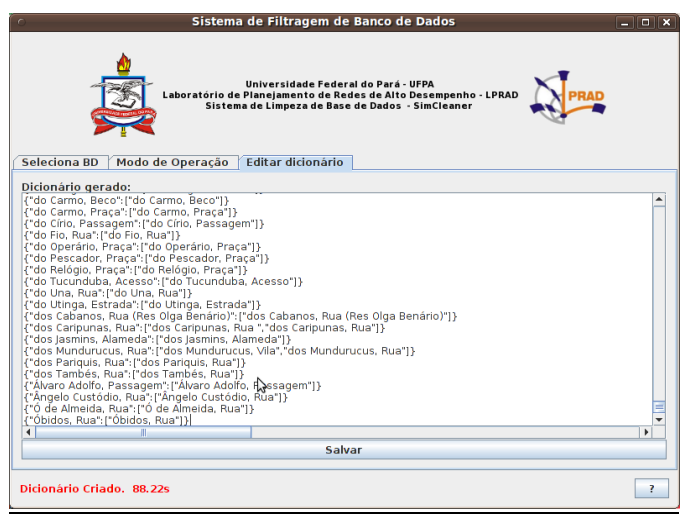

**Figura 2. Edição do dicionário**

O formato escolhido para representar o dicionário foi o padrão *JSON (Java Script Object Notation)*, onde *key* é a chave associada aos valores equivalentes contidos no *array* no lado direito da expressão (**Figura 3**).

{"key": ["value0","value1" ... "valueN"]}

**Figura 3. Representação das palavras no dicionário**

A aplicação carrega o dicionário e itera a tabela realizando as modificações. O processo de filtragem, construção do dicionário e leitura dos arquivos são apresentados em um *log*, que junto do arquivo de saída e do dicionário, são armazenados dentro de um diretório denominado *simcleanerFiles*, inserido no diretório pessoal do usuário.

## 5. Resultados Encontrados

Para ser obtida uma boa analise do comportamento da aplicação, foram executadas dez simulações processando uma quantidade considerável de dados. Os testes foram feitos em um Intel® Core™ i3-540 com 4 Gb de memória RAM e 512 MB de memória swap com o sistema operacional GNU/Linux x86_64 Ubuntu 10.10 com o kernel 2.6.35-25-generic.

**Tabela 1: Desempenho de Criação e Filtragem de arquivos CSV e XLS/XLSX**

|  | Tempo de Criação de Dicionário | | Tempo de Filtragem | |
|---|---|---|---|---|
|  | 3.098 instâncias | 17.782 instancias | 3.098 instâncias | 17.782 instâncias |
| *CSV* | 3.661 s | 49.501 s | 0.042 s | 0.22 s |
| *XLS/XLSX* | 4.052 s | 52.930 s | 0.71 s | 4.717 s |
| *SGBDs* | 87.948 s | 825.848 s | 12.691 s | 31.010 s |

Na **Tabela 1** pode ser observado o desempenho das duas principais funcionalidades da aplicação, onde se evidencia a inferioridade do tempo de processamento de bases no formato *CSV*, isso é creditado ao fato de o *CSV* ser um arquivo de texto puro, enquanto que o *XLSX* é um formato de arquivo compactado, logo sua manipulação demanda um maior uso de memória e processamento para a sua descompactação, no caso dos SGBDs, isso ocorre devido o tempo para acessar os registros do banco de dados ser superior ao de acesso a arquivos comuns.

Foram efetuados testes para comparar o tempo demandado durante a filtragem dos arquivos (**Figura 4**) e criação do dicionário (**Figura 5**) usando CSV, XLSX e dos SGBDs.

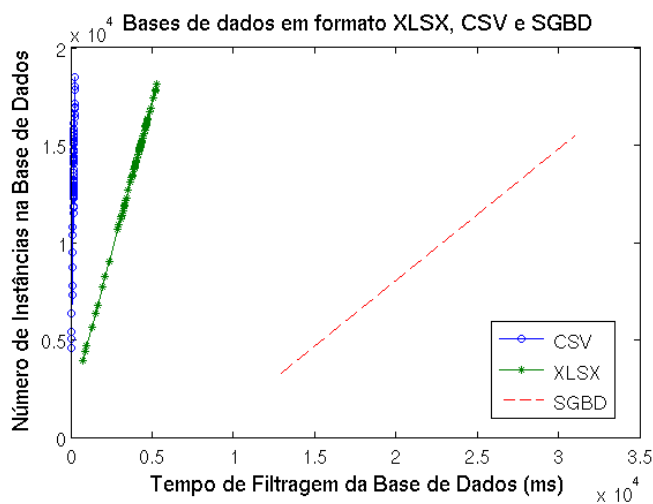

**Figura 4. Filtragem da base.**

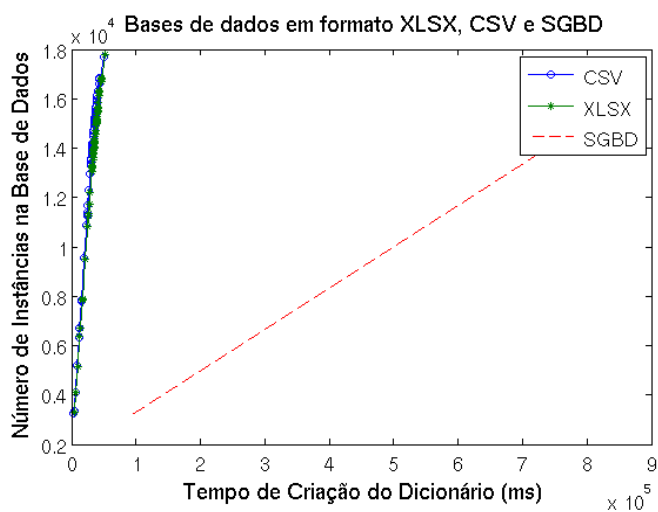

**Figura 5. Criação dos dicionários.**

## 7. Considerações Finais e Trabalhos Futuros

O KDD é um processo que permite a extração de padrões de base de dados. Contudo, o seu sucesso depende da consistência da base. Para obter isso, pode ser necessário um pré-processamento das mesmas antes de serem submetidas ao KDD. O ato de limpar, padronizar e remover ruídos existentes em um BD é chamado de *data cleaning*.

A base de ocorrências policiais da SSP-PA é um exemplo de BD que precisou ser pré-processada para posteriormente ser submetida à mineração de dados. Para solucionar isso, foi proposta uma ferramenta para *data cleaning* usando funções de similaridade baseadas em caracteres. As funções de similaridade baseadas em *tokens* e a comparação do desempenho entre elas estão em um âmbito de trabalhos futuros.

A aplicação é compatível com arquivos CSV, XLS, XLSX e com os SGBDs MySQL e PostgreSQL. Ela se mostrou eficiente no pré-processamento das bases de dados, removendo inconsistências e aumentando a qualidade dos dados, como proposto. Após o uso desta aplicação, técnicas de IC puderam ser aplicadas e resultados mais precisos foram obtidos.

**Referências**